# Race and gender income inequality in the USA: black women vs. white men

Ivan O. Kitov

Institute of Geospheres Dynamics, Russian Academy of Sciences

Abstract
Income inequality between different races in the U.S. is especially large. This difference is even larger when gender is involved. In a complementary study, we have developed a dynamic microeconomic model accurately describing the evolution of male and female incomes since 1930. Here, we extend our analysis and model the disparity between black and white population in the U.S., separately for males and females. Unfortunately, income microdata provided by the U.S. Census Bureau for other races and ethnic groups are not time compatible or too short for modelling purposes. We are forced to constrain our analysis to black and white population, but all principal results can be extrapolated to other races and ethnicities. Our analysis shows that black females and white males are two poles of the overall income inequality. The prediction of income distribution for two extreme cases with one model is the main challenge of this study.

## Introduction

Income inequality is a global phenomenon. It has many dimensions deserving special study. For example, long-term income measurements reveal significant differences between genders, races, and countries. These differences demonstrate strong dependence on age and also evolve with time. Since personal income is a measured variable, one can carry out quantitative analysis and numerical modelling. A correct numerical model should be consistent with data, *i.e.* to accurately predict all characteristics of personal income distribution (PID) and their change over time. This is third in a series of four papers investigating age, gender, race, and country dependence of income distribution. First paper in the series presents our microeconomic model of individual income growth and income distribution for two genders in the U.S. [Kitov and Kitov, 2015a]. It accurately predicts the PID evolution for males and females since 1930. Second paper compares PID characteristics in four countries - the U.S., the UK, Canada, and New Zealand - and finds that the latter three countries repeat the past of the U.S. with time lag defined by the gap in real GDP per capita [Kitov and Kitov, 2015b]. Fourth paper addresses the gender dependence of the difference between countries. In this study, we consider only individual incomes. Households and families aggregate individuals with measured incomes according to various and varying rules. Hence, the household and family income distributions are both exactly derived from the PID, when their structure is known. The measures of income aggregated at fluctuating basis can only complicate quantitative analysis and modelling.

     Here, we study the difference in the age and time evolution of specific features of race and gender dependent personal income. Without loss of generality, two races are selected: white and black. For other races (ethnicities) in the U.S., the Census Bureau does not provide consistent long-term and continuous (compatible in) time series. Definitions of race/ethnicity have been revised several times during the past 50 years. This is the period when income microdata for various races and genders are available from the Integrated Public Use Microdata Series (IPUMS) [King *et al*., 2010]. Since the IPUMS dataset is used in this study to estimate gender and race related characteristics of income the studied period is limited to the years between 1962 and 2014.

     Our analysis includes extensive processing of actual data and theoretical consideration at the level of microeconomic modelling. In contrast to other studies of income inequality, we

quantitatively predict individual incomes in a given economy together with various aggregate income measures using a dynamic model [Kitov and Kitov, 2015a]. This is an evolutionary model, where the change of each and every individual income is driven by one exogenous variable – real GDP per capita estimated for working age population. This model allows to accurately describing all observed characteristics of personal income distribution for all males and all females in the U.S. In this study, we analyse and predict the evolution of two specific age-dependent parameters for four race-gender configurations. These parameters are the mean income and the share of population distributed according to the Pareto law.

The evolution of age-dependent mean income demonstrates the overall lag between income earned by people of different gender and race, which cannot be surmounted without dramatic changes to the root social relationships between races and genders. The second feature highlights the depth of discrepancy in the portion of people of different race and gender with the highest incomes. This disproportion is especially high for the youngest and eldest age categories. The lack of presence in the high-income range makes the shape of the age-dependent mean income distribution for all "not-white-male" combinations to lag by decades behind white-male population. Our microeconomic model reveals the fact that the observed disproportion in gender and race representation in the high-income range has a measurable negative impact on real economic growth in the U.S., not saying about the extremely high level of unjustified social disparity.

In Section 1, we introduce the microeconomic model for personal income distribution and illustrate its performance comparing the observed and predicted mean income and portion of rich people in the U.S. for all males and all females. In Section 2, we analyse the difference between two gender and two races – black and white – in terms of model parameters and formulate principal tasks to model. Essentially, we are looking for features, which were not found, and thus, modelled in the refined model explaining the difference between genders [Kitov and Kitov, 2015a]. Section 3 is devoted to modelling and interpretation.

## 1. Model

We start with a simple graph comparing trajectories of real mean income for two genders and two races – white and black. Left panel of Figure 1 displays the mean personal income for all population with income reported to the Census Bureau [2015a]. Since white population prevails in the U.S. race structure the mean incomes for all male and female are close to those for the white. The black male curve is marginally above the white female one. The dominance of white male, as expressed by mean income, is extraordinary. Is that fair in terms of higher personal capability of white males?

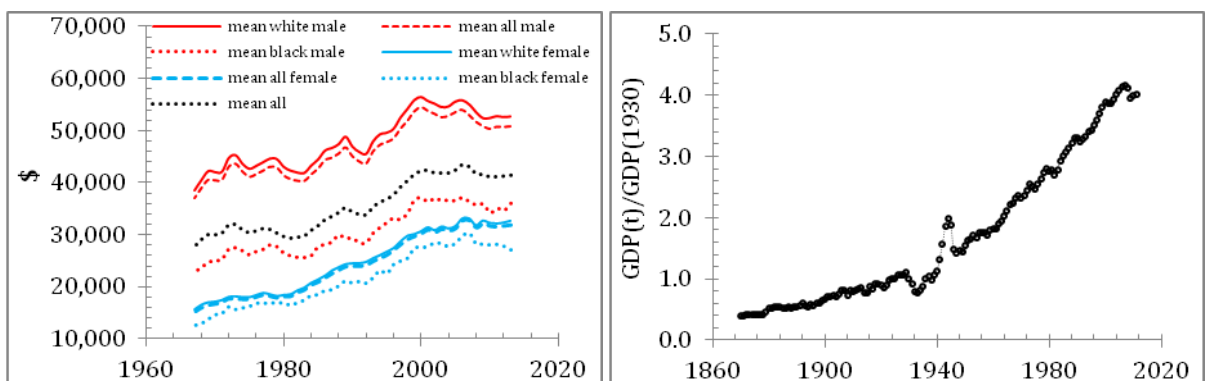

Figure 1. Left panel: The evolution of gender and race (white and black) dependent mean income since 1967 as reported by the U.S. Census Bureau in constant 2013 U.S. dollars. White population prevails in the U.S. race structure, and thus, the mean incomes for all male and female are close to those for the white population. Right panel: The evolution of real GDP per capita from 1870 to 2013. All annual estimates normalized to the 1930 level. Before 1929, the estimates are borrowed from the Maddison dataset.

Following the whole bulk of observations of personal income in the USA as well as in some other developed countries, our gender-dependent microeconomic model [Kitov and Kitov, 2015a] links the change in any personal income, and thus, in the mean income shape with the only parameter – the level of real GDP per capita, *rGDPpc*, as reported by the Bureau of Economic Analysis [2015]. In the right panel of Figure 1, the *rGDPpc* time series is normalized to its level measured in 1930. Before 1929, the real GDP time series is extended by estimates borrowed from the Maddison dataset [Maddison, 2006]. All GDP values after 1900 are corrected for the difference between total and working age population – income is reported only for those who are 15 year old and above. The population estimates are extracted from the Census Bureau population data [2015b]. All in all, we use data only from open sources available for the broader research community.

Figure 2 displays two sets of mean income as a function of age. In the left panel, there are curves for females as measured from 1967 to 2007, and in the right panel similar curves are shown for males in the U.S. The mean income estimates for various genders and races were calculated from individual income data downloaded from the IPUMS, which processes and distributes the March CPS income data. The shapes of mean income curves are different for men and women, and our model explains this fact by smaller work capital available for females. Two most important features in Figure 2 to be modelled are related to clear secular changes in shape. Firstly, the slope of the initial segments of all curves decreases with time, with the male curves having lower slopes than the female ones. Secondly, the age when these curves reach their respective maxima increases with time. And again, the peak age for females is smaller than that for males. Our microeconomic model explains and accurately predicts these and other observed features and their secular changes.

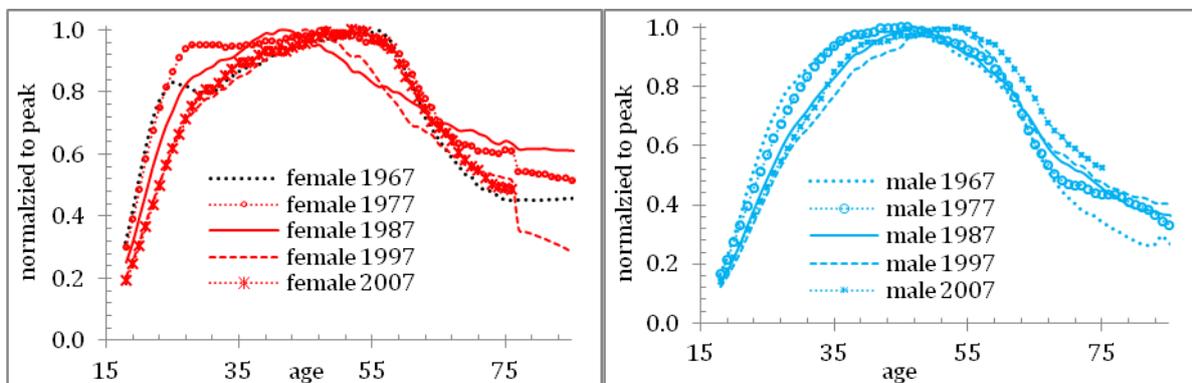

Figure 2. The evolution of observed mean income dependence on age for females (left panel) and males (right panel). All curves are smoothed by MA(7) and then normalized to their peak values.

Gender bisects population, with two approximately equal population subsets revealing quite different properties of income distribution and its evolution. The differences in other personal characteristics like race or ethnicity may be hidden in these larger subsets. Here, we would like to extend the study of males and females to smaller groups representing racial differences. Unfortunately, smaller racial groups are not well presented in the CPS and/or do not create longer times series in the IPUMS due to revisions of race definition or introduction of new racial gradation. We have selected black and white population in order to demonstrate the level of income disparity between races, which is additionally amplified by gender difference. The most prominent difference is expected between white males and black females. It should be in the focus of this study. On the other hand, the characteristics of black males and white females are closer than those for black males and females.

To predict income distribution related to two genders and two races we use our microeconomic model based on physical intuition and observations. In this paper, we give a brief description of the refined model. The complete version is presented in [Kitov and Kitov, 2015a]. To characterize the change in individual income we use the income rate, $M(t)$, i.e. the total income a person earns per year. In essence, $M(t)$ is an equivalent of $Y$ in the Cobb-Douglas production function. The principal driving force of income growth is the personal capability to earn money, $\sigma(t)$, which is similar to labour in the Cobb-Douglas function.

Based on thorough physical consideration of income observations, we introduce a new notation - the rate of dissipation of income and assume that it has to be proportional to the attained level of $M(t)$. The equation defining the change in $M(t)$ should include a term, which is inversely proportional to the size of means or instruments used to earn money (like work capital), which we define by variable $\Lambda(t)$. Then the dissipation term is proportional to $M(t)/\Lambda(t)$. To describe the dynamics of income growth as a function of work experience, $t$, one can write an ordinary differential equation:

$$dM(t)/dt = \sigma(t) - \alpha M(t)/\Lambda(t) \qquad (1)$$

where $M(t)$ is the rate of money income denominated in dollars per year [$/y$], $t$ is the work experience expressed in years [$y$]; $\sigma(t)$ is the capability to earn money, which is a permanent feature of an individual [$\$/y^2$]; $\Lambda(t)$ is the size of the earning means, which is a permanent income source of an individual [$\$/y$]; and $\alpha$ is the dissipation factor [$\$/y^2$].

We assume that $\sigma(t)$ and $\Lambda(t)$ are mutually independent - that is a person's capability to earn money is not related to her work instrument. Notice that we have chosen $t$ to denote the work experience rather than the person's age. All people start with a zero income, $M(0)=0$, which is the initial condition for (1). At $t=0$, all incomes are zero, and then they start to change according to (1) as $t>0$. Notice that both $\sigma(t)$ and $\Lambda(t)$ vary with $t$. This means that (1) has to be solved numerically, which is the approach we apply to calibrate the model to data. Before proceeding to the calibration stage, we first make a few simplifying assumptions, under which the model has a closed-form solution.

We introduce a modified capability to earn money:

$$\Sigma(t) = \sigma(t)/\alpha \qquad (2)$$

To distinguish between cohorts, we introduce absolute time flow, $\tau$. The time flow for work experience, $t$, and calendar years, $\tau$, relate to each other in a natural fashion. Variables $\Lambda$ and $\Sigma$ depend on $\tau$, thereby introducing differences in income capability and instrument among age cohorts. Our model captures cross sectional and intertemporal variations. In line with the Cobb-Douglas production function, we make a simplifying assumption by letting $\Lambda(\tau_0,t)$ and $\Sigma(\tau_0,t)$ to evolve as the square root of the increment in the aggregate output per capita. The capability and instrument evolve according to:

$$\Sigma(\tau,t) = \Sigma(\tau_0,t_0) [Y(\tau)/Y(\tau_0)]^{1/2} \qquad (3)$$

$$\Lambda(\tau,t) = \Lambda(\tau_0,t_0) [Y(\tau)/Y(\tau_0)]^{1/2} \qquad (4)$$

where $t=\tau-\tau_0$, $\Sigma(\tau_0,t_0)$ and $\Lambda(\tau_0,t_0)$ are the initial values of capability and instrument for a person with zero work experience in year $\tau_0$; $Y(\tau_0)$ and $Y(\tau)$ are the aggregate output per capita values in the years $\tau_0$ and $\tau$, respectively, and $dY(\tau_0,t)=Y(\tau)/Y(\tau_0)=Y(\tau_0+t)/Y(\tau_0)$ is the cumulative output growth. Now we restrict our attention to the initial values of the capability and instrument as functions of the initial year: $\Lambda(\tau_0)$ and $\Sigma(\tau_0)$, respectively. The product of equations (3) and (4), $\Sigma(\tau_0,t_0)\Lambda(\tau_0,t_0)$, evolves with time in line with growth of real GDP per capita as in the Cobb-Douglas production function. We call $\Sigma\Lambda$ the capacity to earn money, which means that $\Lambda(\tau_0,t_0)\Sigma(\tau_0,t_0)$ is the initial capacity.

Equation (1) can be re-written to account for the dependence on the initial year, $\tau_0$:

$$dM(\tau_0,t) / dt = \alpha\{\Sigma(\tau_0,t) - M(\tau_0,t) / \Lambda(\tau_0,t)\} \qquad (5)$$

Personal incomes have lower and upper limits. We assume that the capability to earn money, $\Sigma(\tau_0,t)$, and the size of earning means, $\Lambda(\tau_0,t)$, are also bounded above and below. Then they have positive minimum values among all persons, $k = 1, \ldots, N$, with the same work experience $t$ in a given year $\tau_0$: $\min\Sigma_k(\tau_0,t)=\Sigma_{min}(\tau_0,t)$ and $\min\Lambda_k(\tau_0,t)=\Lambda_{min}(\tau_0,t)$, respectively, where $\Sigma_k(\tau_0,t)$ and $\Lambda_k(\tau_0,t)$ are the parameters corresponding to a given individual. We can formally introduce the relative and dimensionless values of the defining variables in the following way:

$$S_k(\tau_0,t) = \Sigma_k(\tau_0,t) / \Sigma_{min}(\tau_0,t) \qquad (6)$$

and

$$L_k(\tau_0,t) = \Lambda_k(\tau_0, t) / \Lambda_{min}(\tau_0,t) \qquad (7)$$

where $S_k(\tau_0,t)$ and $L_k(\tau_0,t)$ are the dimensionless capability and size of work instrument, respectively, for the person $k$, which are measured in units of their minimum values. So far, all $N$ persons in the economy are different and at this stage of model development we need to introduce proper distributions of $S_k(\tau,t)$ and $L_k(\tau,t)$ over population.

We have justified the choice of discrete uniform distributions for $S_k$ and $L_k$ [Kitov 2005b]. The relative values of $S_k(\tau_0,t_0)$ and $L_k(\tau_0,t_0)$, for any $\tau_0$ and $t_0$, can have only discrete values from 2 to 30. There are 29 values of $S_i(\tau_0,t_0)$ and $L_j(\tau_0,t_0)$: $S_1(\tau_0,t_0)=2, \ldots, S_{29}(\tau_0,t_0)=30$, and similarly for $L_j(\tau_0,t_0)$, where $j=1,...,29$. We assume a uniform distribution between 29 different capabilities as the simplest one, and thus, the entire working age population is divided into 29 equal groups. All $k$ work instruments are uniformly distributed over 29 different sizes from 2 to 30 as well.

The probability for a person to get an earning means of relative size $L_j$ is constant over all 29 discrete values of the size and the same is valid for $S_i$. The distribution over income involves the history of work experience $t$ described by (1), and thus, differs from the distribution over relative values. The relative capacity for a person to earn money is distributed over the working age population as the product of the independently distributed $S_i$ and $L_j$:

$$S_i(\tau,t)L_j(\tau,t) = \{2\times2, \ldots, 2\times30, 3\times2, \ldots, 3\times30, \ldots, 30\times30\}$$

There are 29×29=841 different values of the normalized capacities available between 4 and 900. Some of these cases seem to be degenerate (for example, 2×30=3×20=4×15=...=30×2). However, $\Sigma$ and $\Lambda$ have different influence on income growth in (1) and each of 841 $S_iL_j$ combinations define a unique time history.

Our model contains time varying parameters, and thus, numerical methods have to be applied to solve it and to calibrate to data. However, some useful estimates of principal characteristics can be done in a simplified case, when $\Sigma(\tau_0,t)$ and $\Lambda(\tau_0,t)$ are constant over $t$.

Given constant $\Sigma$ and $\Lambda$, as well as the initial condition $M(0)=0$, the general solution of equation (4) is as follows:

$$M(t) = \Lambda\Sigma[1 - \exp(-\alpha t/\Lambda)] \qquad (8)$$

One can re-arrange equation (8) to dimensionless form using relative measures of income. Ultimately, we are interested in relative income distribution which characterizes income inequality. We first substitute in the product of the relative values $S_i$ and $L_j$ and the time dependent minimum values $\Sigma_{min}$ and $\Lambda_{min}$ for $\Sigma$ and $\Lambda$, and then normalize the equation to the maximum values $\Sigma_{max}$ and $\Lambda_{max}$ in a given calendar year, $\tau$, for a given work experience, $t$. The normalized equation for the rate of income, $M_{ij}(t)$, of a person with capability, $S_i$, and the size of earning means, $L_j$, where $i, j = \{2, \ldots, 30\}$ is as follows:

$$M_{ij}(t) / [S_{max}L_{max}] = \Sigma_{min}\Lambda_{min}(S_i/S_{max})(L_j/L_{max})\{1 - \exp(-\alpha t/[(\Lambda_{min}L_{max})(L_j/L_{max})])\} \qquad (9)$$

or compact:

$$\tilde{M}_{ij}(t) = \Sigma_{min}\Lambda_{min}\tilde{S}_i\tilde{L}_j[1 - \exp(-t(1/\Lambda_{min})(\tilde{\alpha}/\tilde{L}_j))] \qquad (10)$$

where

$$\tilde{M}_{ij}(t) = M_{ij}(t)/(S_{max}L_{max}); \quad \tilde{S}_i = S_i/S_{max}; \quad \tilde{L}_j = L_j/L_{max}; \quad \tilde{\alpha} = \alpha/L_{max}$$

and $S_{max}=L_{max}=30$.

For constant $L_j$ and $S_i$, one can derive from (10) the time needed to reach the absolute income level H, where H<1:

$$t_H = \Lambda_j\ln[1-H)] / \alpha \qquad (11)$$

This equation is correct only for persons capable to reach H, i.e. when $L_jS_i/S_{max}L_{max}>H$. With all other terms in (11) being constant, the size of work instrument available for a person, $\Lambda_j$, defines the change in $t_H$. In the long-run, $t_H$ increases proportionally to the square root of the real GDP per capita.

For the initial segment of income growth, when $t<<1$, the term $\alpha t/\Lambda$ in (11) is also << 1. One can derive an approximate relationship for income growth by representing the exponential function as a Taylor series and retaining only two first terms. Then (8) can be re-written as:

$$M_{ij}(t) = \Sigma_i\Lambda_j \alpha t/\Lambda_j = \Sigma_i\alpha t \qquad (12)$$

i.e. the money income, $M$, for a given person is a linear function of time since $\Sigma_i$ and $\alpha$ are both constants.

The growth trajectory of income described by equation (1) does not present the full picture of income evolution with age. As numerous empirical observations show (*e.g.*, right panel in Figure 2), the average income reaches its peak at some age and then starts declining. In our model, the effect of exponential fall is naturally achieved by setting the money earning capability $\Sigma(t)$ to zero at some critical work experience, $t=T_c$.

The solution of (1) for $t>T_c$ then becomes:

$$\tilde{M}_{ij}(t) = \tilde{M}_{ij}(T_c)\exp[-(1/\Lambda_{min})(\tilde{\gamma}/\tilde{L}_j)(t - T_c)] \qquad (13)$$

and by substituting (9) we can write the following decaying income trajectories for $t>T_c$:

$$\tilde{M}_{ij}(t) = \Sigma_{min}\Lambda_{min}\tilde{S}_i\tilde{L}_j\{1 - \exp(-(1/\Lambda_{min})(\tilde{\alpha}/\tilde{L}_j)T_c)]\exp\{-(1/\Lambda_{min})(\tilde{\gamma}/\tilde{L}_j)(t - T_c)\} \quad (14)$$

First term in (14) is the level of income rate attained at $T_c$. Second term expresses the observed exponential decay of the income rate for work experience above $T_c$. The exponent index $\tilde{\gamma}$ represents the rate of income decay that varies over time and is different from $\tilde{\alpha}$. It was shown in Kitov [2005a] that the exponential decay of personal income rate above $T_c$ results in approximately the same relative level at the same age, when normalized to the maximum income for this calendar year. This means that the decay exponent can be obtained according to the following relationship:

$$\tilde{\gamma}(t) = -\ln A / [T_A - T_c(t)] \quad (15)$$

where $A$ is the constant relative level of income rate at age $T_A$. Thus, when the current age reaches $T_A$, the maximum possible income rate $\tilde{M}_{ij}$ (for $i = 29$ and $j = 29$) drops to $A$. Income rates for other values of $i$ and $j$ are defined by (14).

In the original model, relationships (14) and (15) were applied to all incomes $M_{ij}$. In the refined model, both equations work only for incomes in the range of Pareto distribution. This modification was introduced in order to distinguish between the male and female trajectories of age-dependent mean income. For females, the input of rich people, i.e. those whose incomes are distributed by a power law, in the 1960s and 1970s was negligible. The shape of 1967 and 1977 female curves in the left panel of Figure 2 does not show any sharp peak between 20 and 60 years of age, while the male curves in the right panel demonstrate consistent growth in $T_c$. Therefore, incomes in the low and middle ranges, like all women's incomes in the 1960s, are not affected by $T_c$. However, they have a different critical age, as discussed below.

The critical age in (13-14) is not constant. For example, the male curves in the right panel of Figure 2 demonstrate that $T_c$ has been increasing between 1962 and 2011. Therefore, its dependence on the driving force of income distribution - real GDP per capita - has to be one of central elements of our dynamic model since any model should match the long-term observations. To predict the increase in $T_c(\tau)$ we use (11). Assuming that the peak value of the mean income is constant in relative terms, we obtain:

$$T_c(\tau) = T_c(\tau_0) [Y(\tau) / Y(\tau_0)]^{1/2} \quad (16)$$

This was the setting in the original model. But there are two features in the females' curves in Figure 2, which likely manifest some changes in the parameters considered in the original setting as constant. In Figure 2, we observe a convergence trend of the male and female curves since the earlier 1960s. Within our framework, this trend can be explained by the change in size of work instruments available for women in addition to that induced by GDP growth. In the original model, we used fixed relative sizes of work instruments for all people. In the refined version, our model includes an option allowing the original instrument size for a given gender, $\Lambda_j$, to change as multiplied by external factor FL($t$). For all females, FL changes from 0.45 in 1962 to 0.65 in 2014 as linear function of time [Kitov and Kitov, 2015a]. Different time (or GDP) dependences are also possible, but they have to be justified by data, which fluctuate too much to provide precise estimates for more complicated functions.

A lower FL in the 1960s and 1970s provides a lengthy "no-change" period before the critical age, as observed in the left panel of Figure 2. Moreover, the critical age for women observed between 1962 and the earlier 1980s is almost constant and close to the retirement age. It is reasonable to suggest that there are two critical ages – one for the highest incomes, $T_c$, and another for the low-middle incomes, $T_S$. The former variable depends on GDP. The latter one – is practically constant and may correlate with retirement age. It is easy to model the retirement effect by an exponential fall in the portion of people in the labour force [Munell, 2011]. Using the basic concept of (15) one can write the following equation:

$$\eta = -\ln B / (T_B - T_S) \tag{17}$$

where $\eta$ is the exponential index, $B$ is the constant relative level of income rate at the age $T_B > T_S$. Both constants in (17) have to be estimated from data. The exponent $\eta$ does not depend on time or real GDP per capita.

The initial exponential (saturation) growth and following fall-off, however, do not complete our model, where persons with the highest $S$ and $L$ may have income only by a factor of 225 larger than that received by persons with the smallest $S$ and $L$. The actual ratio of the highest and lowest incomes is tens of millions, if to consider the smallest reported of $1. Moreover, the defining equation for income growth is not able to predict a power law distribution. We still need to introduce special treatment for the very top incomes that in multiple empirical studies have been shown to follow the Pareto distribution.

Fortunately, it is not necessary to quantitatively predict the distribution of the highest incomes. We adapt to our income model a concept distinguishing the below-threshold (sub-critical) and the above-threshold (super-critical) behaviour of earners. The dynamics of the system in the sub-critical state (low and middle incomes) is described by the equations (1-17). In the super-critical regime, the frequency distribution of incomes is described by a power law. The sub-critical dynamics exactly predicts the portion of people in the super-critical state and the age of transition.

In the model we introduce a critical level of income separating two income regimes: the quasi-exponential (sub-critical) and the Pareto one (super-critical). We call this level "the Pareto threshold", $M^P(\tau)$. Below this threshold, in the sub-critical income zone, personal income distribution (PID) is accurately predicted by the model for the evolution of individual incomes. Above the Pareto threshold, in the super-critical income zone, the observed PID is best approximated by an empirically estimated power law. The excellent predictive power of our model is determined by the possibility to accurately describe the dependence of the portion of people above $M^P$ on age as well as the evolution of this dependence over time.

In the original model, the estimated dimensionless Pareto threshold for all people disregarding gender and race is $M^P(\tau_0)=0.43$ [Kitov, 2005a]. Without loss of generality, we can define the initial real GDP per capita as 1. In this case, $M^P(\tau_0)=0.43$ for any starting year, where $Y(\tau_0)=1$. To match the change in incomes related to real economic growth the absolute value of the Pareto threshold evolves with time proportionally to growth in real output per capita:

$$M^P(\tau) = M^P(\tau_0) [Y(\tau) / Y(\tau_0)] = M^P(\tau_0) Y(\tau) \tag{18}$$

This retains the portion above this threshold almost constant over time.

We found that the Pareto threshold is different for males and females [Kitov and Kitov, 2015a]. This difference is one of principal features of the gender disparity in the U.S. and deserves deeper analysis and special modelling as applied to various race-gender combinations. For all females, the Pareto threshold is linked to FL($t$), and thus, follows a linear time trend from its initial value of 0.29 in 1960 to 0.39 in 2014. The linear time dependence can be changed according to the FL time dependence. All in all, it was easier to insert the observed lower Pareto threshold for women in our model than to understand the forces behind such a difference.

Figure 3 depicts the observed and predicted number of females and males above their specific Pareto thresholds in 1962 and 2011. The shape match is excellent: the age when predicted and observed curves start to grow, peak values, the age of peak values, the rate of fall, etc. Moreover, the model predicts the change in the number of people from 1962 to 2011. Therefore, the model predicts PIDs for each gender and each year together with their dynamic change over 50 years as a sole function of GDP per capita. The most striking difference is between males and females in 1962 - there are 40,000 women 150,000 men in the peak age cell. Notice that the Pareto threshold in 1962 was 0.29 for females and 0.43 for males. The total portion of females above the male Pareto distribution is by a factor of 100 lower than the males' portion. As discussed above, the influence of rich women on mean income was negligible in 1962.

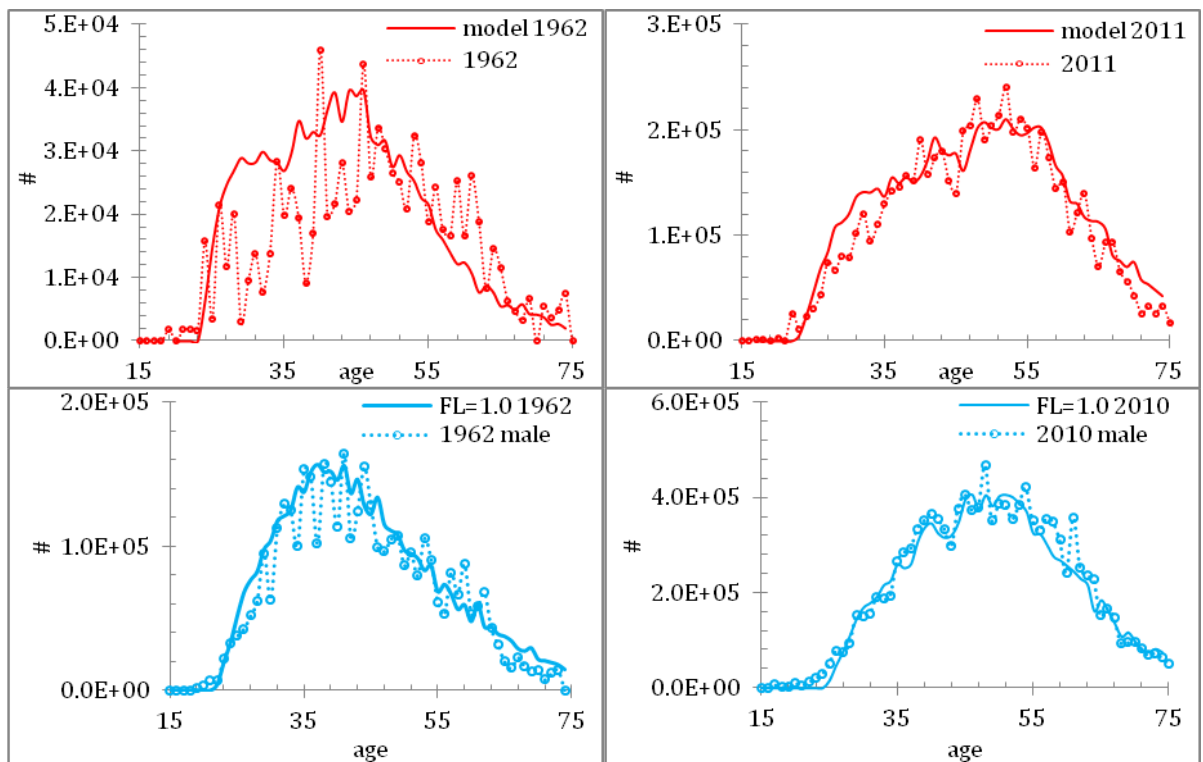

Figure 3. The portion of people above the Pareto threshold as a function of age for females (upper panels) and males (low panels) for 1962 (left) and 2011/2010 (right).

The refined model accommodates all principal observations for two genders and includes all necessary parameters to describe the distribution of low, middle, and top incomes, whatever are their sources and changes over time. Figure 4 illustrates the predictive power of the refined microeconomic model of income growth and distribution by comparison of the predicted and measured mean incomes in 1975 for females and in 1992 for males. The match is especially good before the critical age, $T_c$. This is the period when all defining parameters, except those related to the exponential fall, and underlying distributions can be most accurately estimated. The segment between $T_c$ and the age of retirement for males demonstrates the importance of equations (15) and (17). Beyond $T_c$, all defining equations for sub-critical range are also applicable and allow us to predict the number of people above the Pareto threshold. Overall, the mean income curve aggregates all phenomena described in equations (1)-(18) together with the uniform distribution of personal capabilities and sizes of instruments.

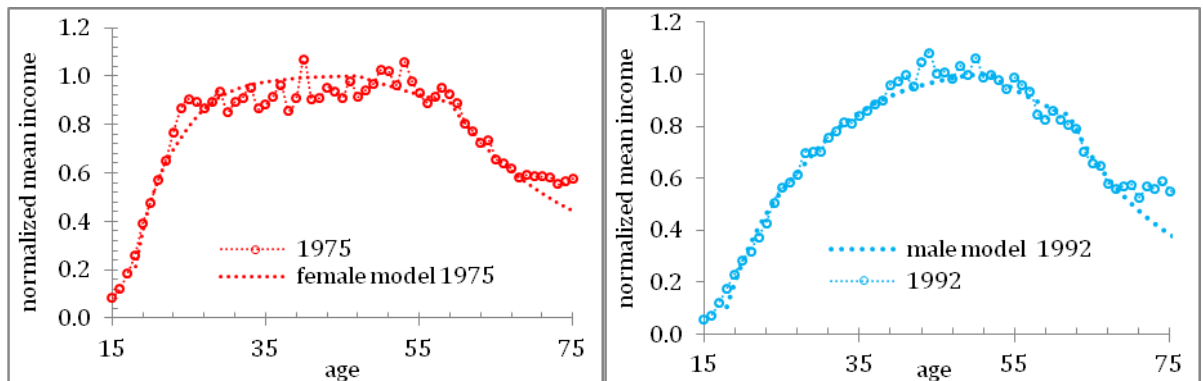

Figure 4. The observed and predicted evolution of mean income. Left panel: The actual mean income estimates for females (red circles) from the IPUMS income data. The predicted curve (dashed line) contains three segments: exponential saturation curve before critical time $T_c$, exponential fall between $T_c$ and the retirement age $T_R$, and double exponential fall beyond $T_R$. Right panel: Same as in the left panel for males in 1992.

The male and female curves in Figures 2 and 3 do converge with time and the refined model has to catch this convergence using internal or external parameters. In the current version, the time-dependent factor FL($t$) serves as an additional degree of freedom to fit observations for all females. We do not need to introduce new parameters controlling the dynamics of income evolution. The male-female income convergence has another aspect, however. Since the female curves converge to the male ones one can imagine that there was time when the male-female difference was even larger than in 1962. The IPUMS data are limited to this year and we cannot figure out what was before 1962. The historical sets of CPS data between 1947 and 1961 are aggregated in ten-year age cells prohibiting accurate estimates of the involved parameters.

The black female mean income curve in Figure 1 suggests that this combination of race and gender lags behind all other combinations in terms of income distribution. Taking into account the dependence of all income features on real GDP per capita one can guess that the features observed for black females in 1962 may repeat those observed in overall female population a few years before. Hence, analysis of black women income distribution may reveal some new features not seen in the overall female curves. Then the refined model has to be modified accordingly. Our study has two objectives - to find new features and to model them within the same framework. Section 2 is devoted to observations.

## 2. Race and gender: income distribution

The difference in nominal mean income as a function of age is illustrated in Figure 5. We compare two genders and two races – black (race_lbl=200 in the IPUMS dataset) and white (race_lbl=100). The income microdata we use in this study allow averaging in one-year of age cells. All mean incomes in Figure 5 are measured in current US dollars. In the left panel, four curves for 1962 are shown for two races (white – W; black –B) and two genders (females – F; males – M). The white-male (WM) curve is far above all other curves almost everywhere except a short period after work start, i.e. work experience less than 5 years. The white-female (WF) curve is slightly below the black-male (BM) curve, and the black-female (BF) curve is the lowermost. In 2014 (right panel), the situation is formally the same but the gap between four curves is much smaller –the BM, WF, and BF curves are very close. The white males' domination in income distribution is still unchallengeable, however. The changing difference between male and female mean incomes has been modelled in a separate study [Kitov and Kitov, 2015]. Here, we analyse the race-specific difference between gender-specific characteristics. We do not compare aggregate features of two races.

The estimates in one-year cells are subject to high-amplitude fluctuations. Since the black population is a just small portion of the white population, the amplitude is much higher for blacks due to lower representation in the CPS population universe. This is a common feature for small groups, which makes modelling and statistical assessment less reliable. The increasing cell width may suppress these fluctuations but could also introduce a significant uncertainty in the estimates of peak age and the age of retirement discussed in Section 1. We apply a centred moving average (MA) to smooth data. This procedure is a compromise between higher fluctuations and larger time uncertainty.

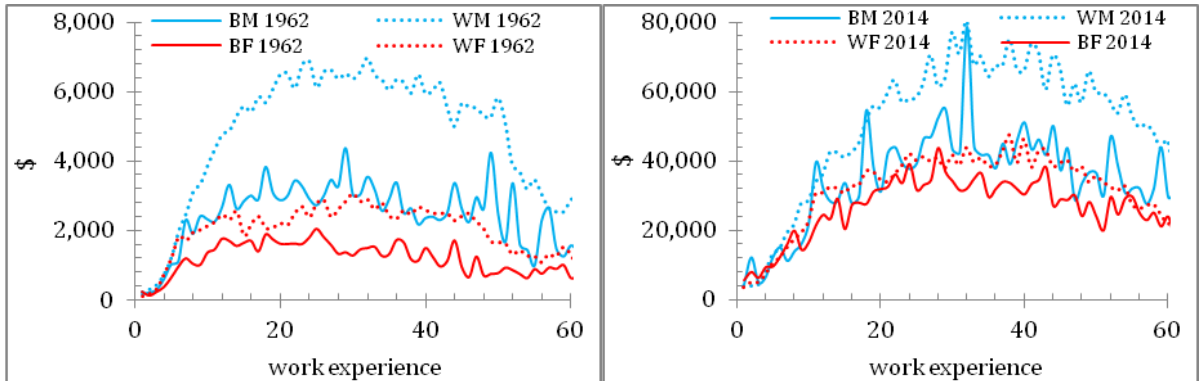

Figure 5. Comparison of age-dependent mean income curves for two races (white – W; black –B) and two genders (females – F; males – M): 1962 (left panel) and 2014 (right panel).

Figure 6 demonstrates the evolution of nominal mean income with time for all four gender-race combinations. Both female curves reveal periods of constant mean income in 1967 and 1977. Since 1987, there are peaks in the female curves illustrating the growth in critical age, which is difficult to estimate, however. All cases are characterized by increasing amplitude of fluctuations which may hide the evolution of critical age $T_c$. The BM curve also has a shelf in 1967 and 1987. This feature is likely linked to the closeness of the BM and WF mean income curves. The relative size of work capital for all groups is smaller than that for WM.

The amplitude of all curves increases with time as a result of nominal growth in GDP per capita. In order to directly compare mean income curves for different years we have smoothed them with MA(7) and then normalized to their respective peak values. Figure 7 depicts the normalized curves for four cases. The WM and WF curves are similar to those for all males and all females, respectively. This is a consequence of dominance of white population in U.S. race structure. The overall mean income curves for males and females were analysed and modelled in a different study [Kitov and Kitov, 2015]. Therefore, their features were accurately predicted by the refined model.

The black population may reveal some new features, which were not covered by our model. This is the primary purpose of race-related data analysis to find such features. When all features for all races are similar no model refinement is needed beyond that carried out for two genders. The BF curves demonstrate the largest shape changes with time, which are likely beyond those observed for all females. The 1967 BF curve peaks at the age around 30 and then falls along a linear trend to the age of retirement. After the retirement age, the curve falls exponentially. The 1977 curve demonstrates general move to the shape of a similar WF curve, i.e. fast growth to some level, which is then retained to the age of retirement. In a sense, the 1977 and 1987 curves are transient cases to the mean income function observed in 2007, which is similar to the corresponding WM curve. The transition from 1967 to 2007 was not observed before and is likely the challenge to our model. To understand better this transition we need more information on the highest incomes for these years.

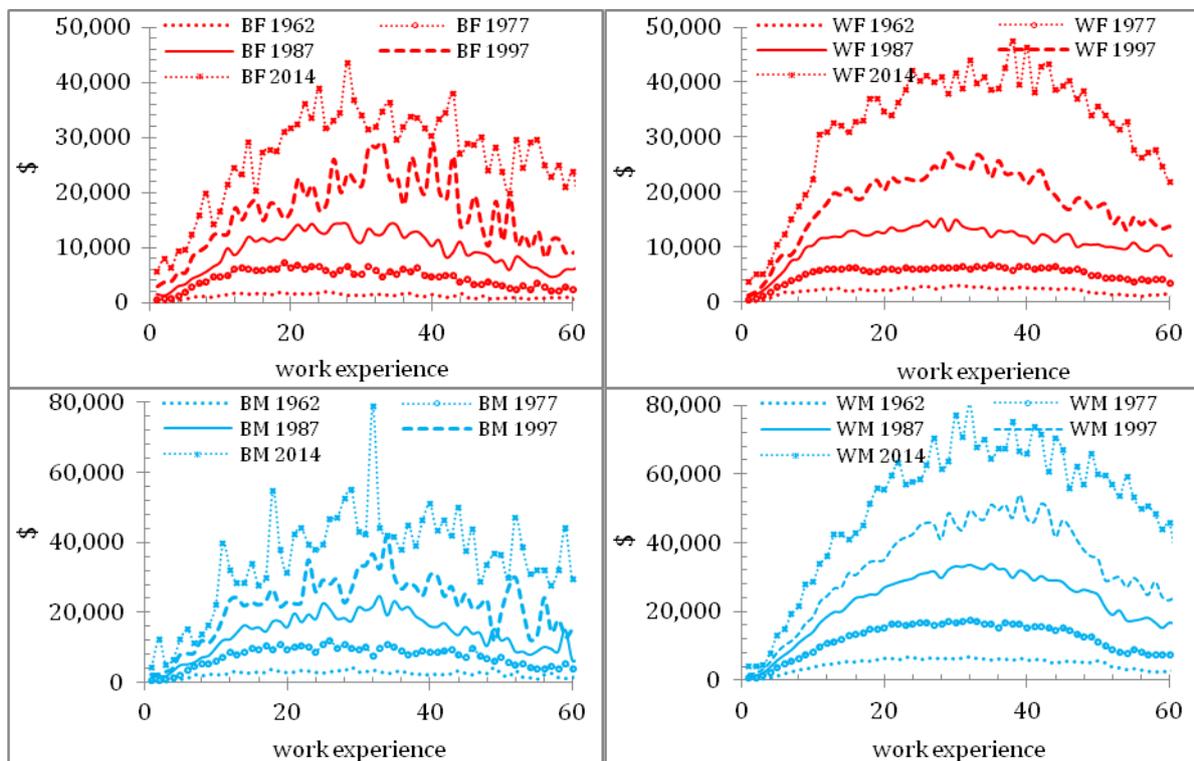

Figure 6. Comparison of nominal mean incomes for two races (white – W; black –B) and two genders (females – F; males – M) between 1962 and 2014. The annual estimates for the years after 1977 are characterized by higher fluctuations. Smoothing is needed.

The curve observed in 1997 is unusual for all cases except WM. For black men and women, the growth rate of 1997 curve is lower than for other years. However, the initial segments of the 1997 curves do not differ much from those observed in 1987 and 2007. As we discussed in Section 1, the initial growth is the most dynamic part of the mean income curve and is directly related to the absolute size of work instrument, and thus, to the level of real GDP per capita. The divergent character of the 1997 curve may be related to real changes or reflect the low quality of corresponding measurements. The latter case seems to be a better explanation in view of more detailed observations depicted in Figure 8.

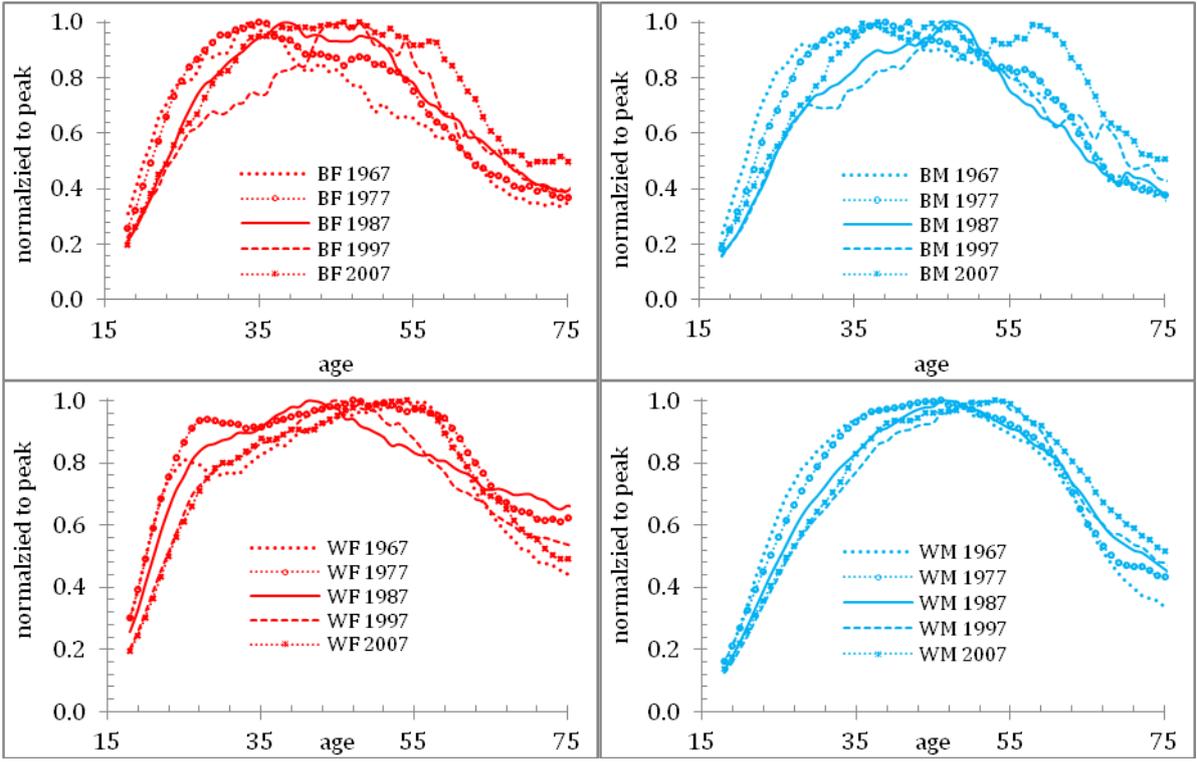

Figure 7. The evolution of age-dependent mean income for two races and two genders between 1967 and 2007. BM – black males, BF – black females, WM – white males, WF – white females.

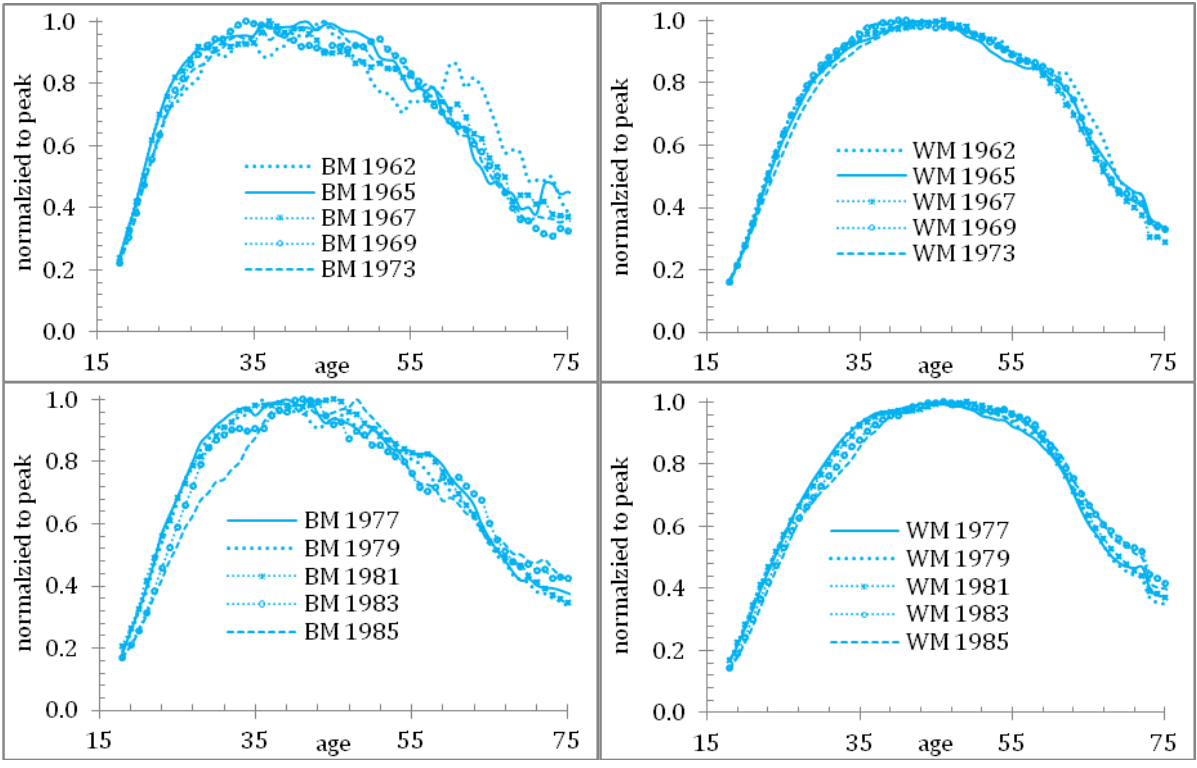

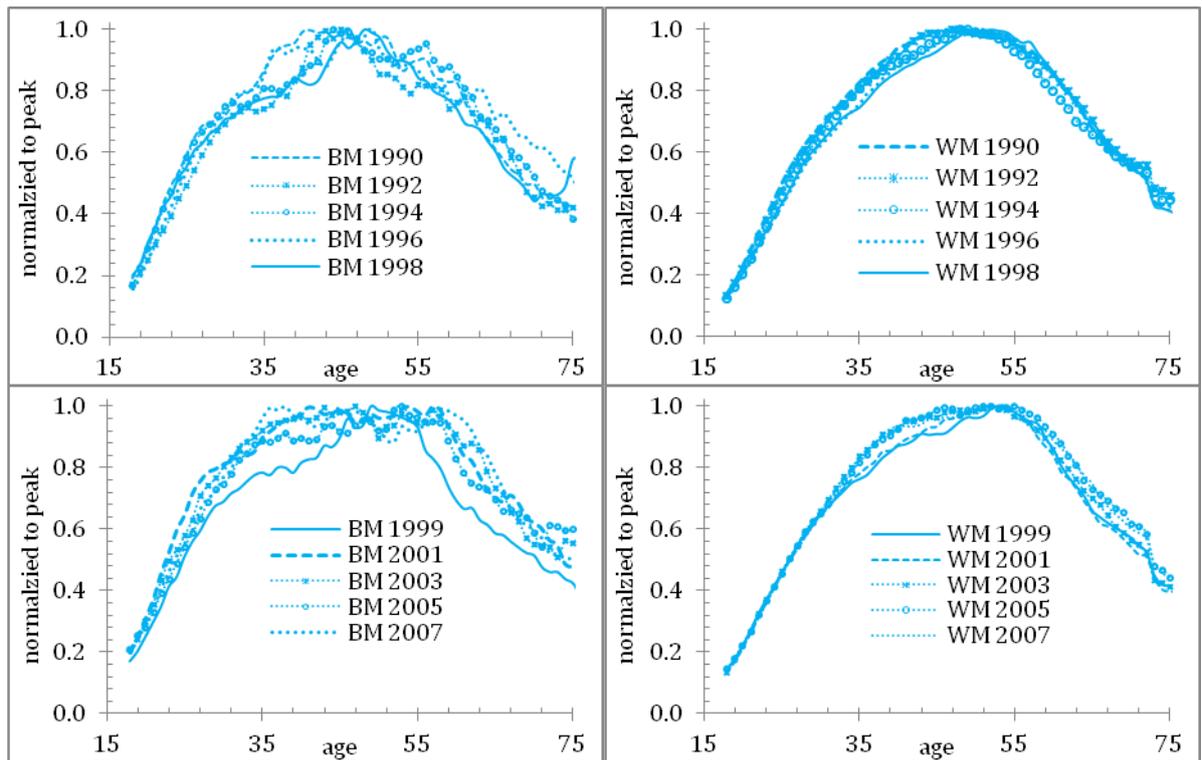

Figure 8. Comparison of black (BM) and white (WM) males. The original mean income curves with annual estimates were smoothed with MA(7) and then normalized to their peak values. The estimates for WM are characterized by lower fluctuations directly related to better coverage. The evolution of WM and BM mean income is different.

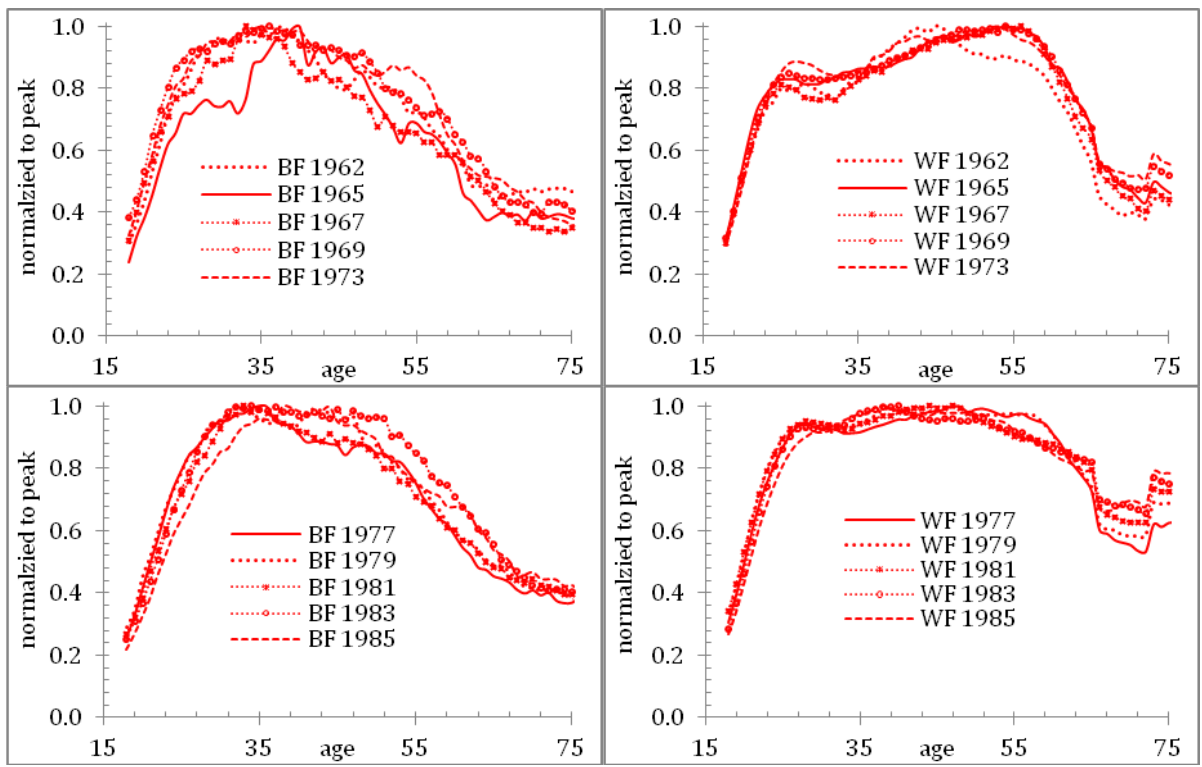

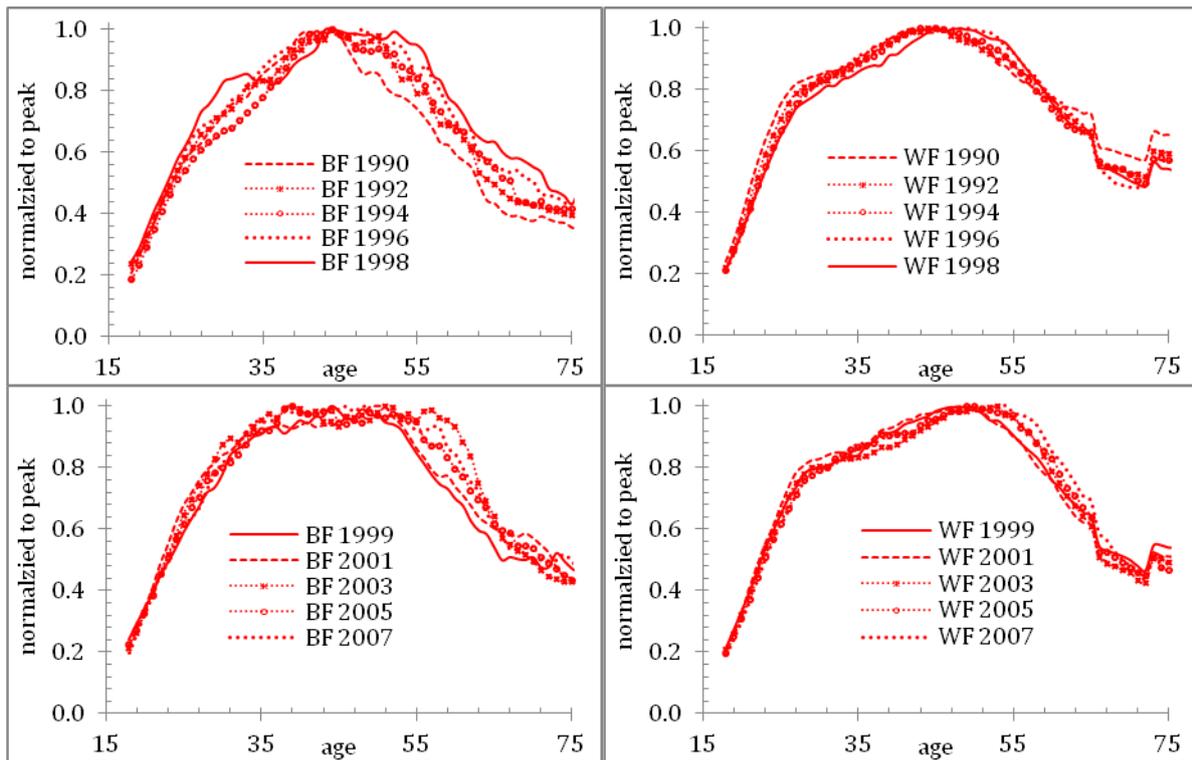

Figure 9. Comparison of black (BF) and white (WF) females. The original mean income curves with annual estimates were smoothed with MA(7) and then normalized to their peak values. The estimates for WF are characterized by lower fluctuations directly related to better coverage. The evolution of WF and BF mean income is different.

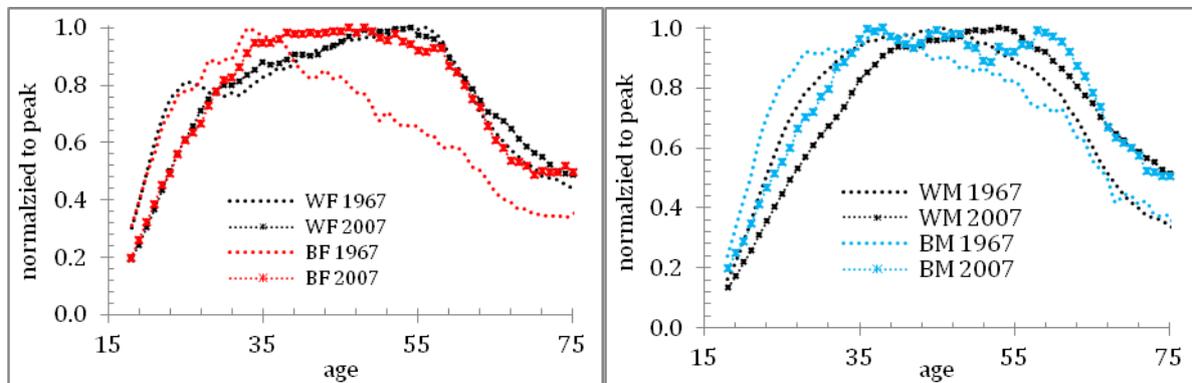

Figure 10. Comparison of age-dependent mean income curves for black males and white females (left panel) and white males and black females.

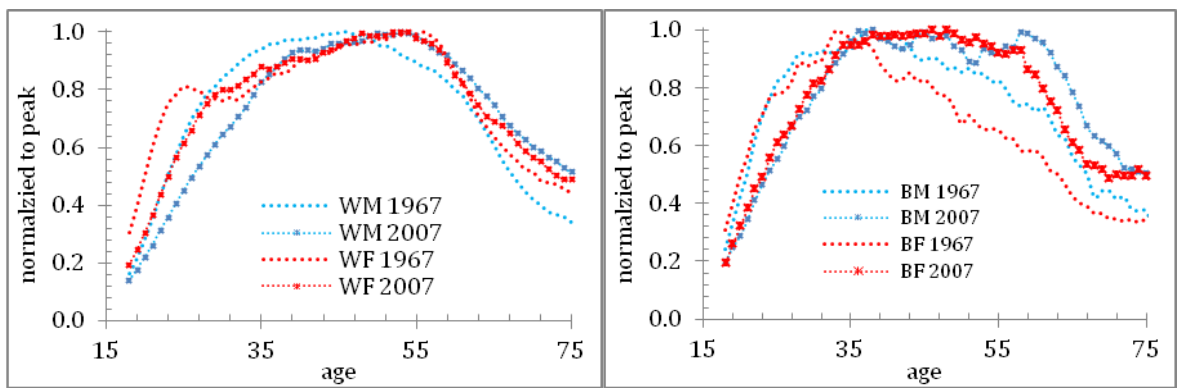

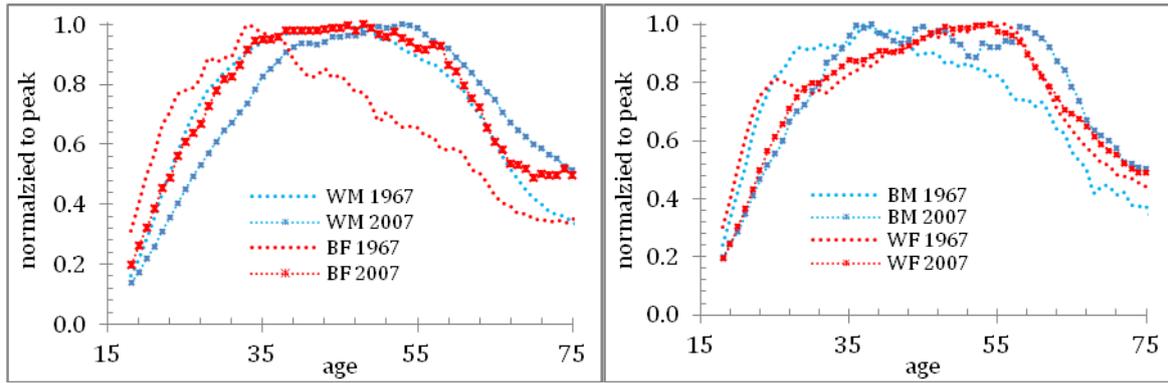

Figure 11. Comparison of the age-dependent mean income for white male and black female (left panel) and black male and white female (right) panel.

The portion of rich people above the Pareto threshold demonstrates strong variations between genders, races, ages, and over time. Figure 12 illustrates the evolution of the portion of rich people from 1967 to 2007. The disparity of black population is striking – the share of black women in 1967 having incomes above $11,000 is next to zero - just a few random non-zero measurements. At the same time, the WM portion is approximately 30% for ages between 40 and 50. The BM and WF shares both hover between 1% and 4%, but the WF is slightly lower in line with the lower mean income figures.

The BF share rises above 5% for ages between 40 and 55 only in 2001 and continues to grow in 2014. A more robust growth in observed in BM and WF data, the latter demonstrate lower fluctuations because of better population coverage, as discussed for the mean income. The joint growth in BF, WM, and WF does not add rich people. The WM population is rather displaced from the top income range to the low-middle income. The portion of WM falls closer to 20% in 2014. The secular fall in the WM share suggests that they are not better than other people in terms of their capability to earn money. Projection of the equalizing trend between genders and races into the future implies that the peak portion of rich people has to be around 28%. The age of the peak share is growing as the root square of real GDP per capita. On average, rich people are getting older and older independent on their race and gender.

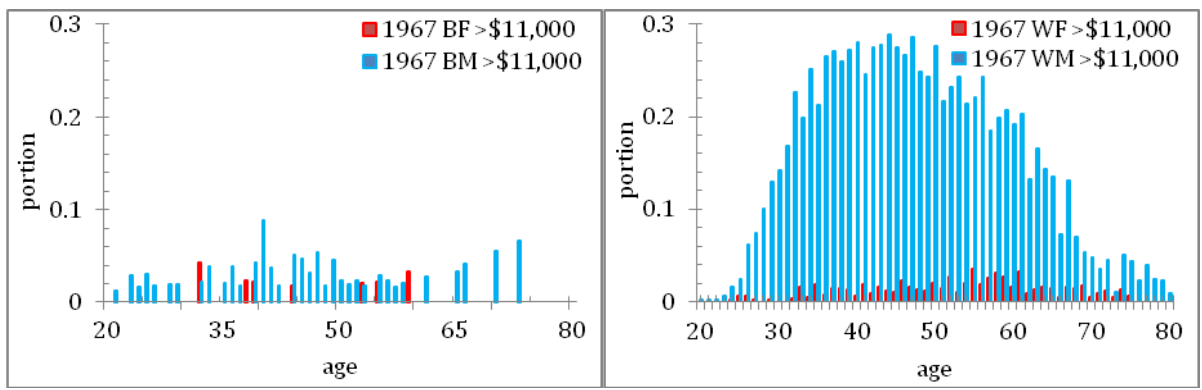

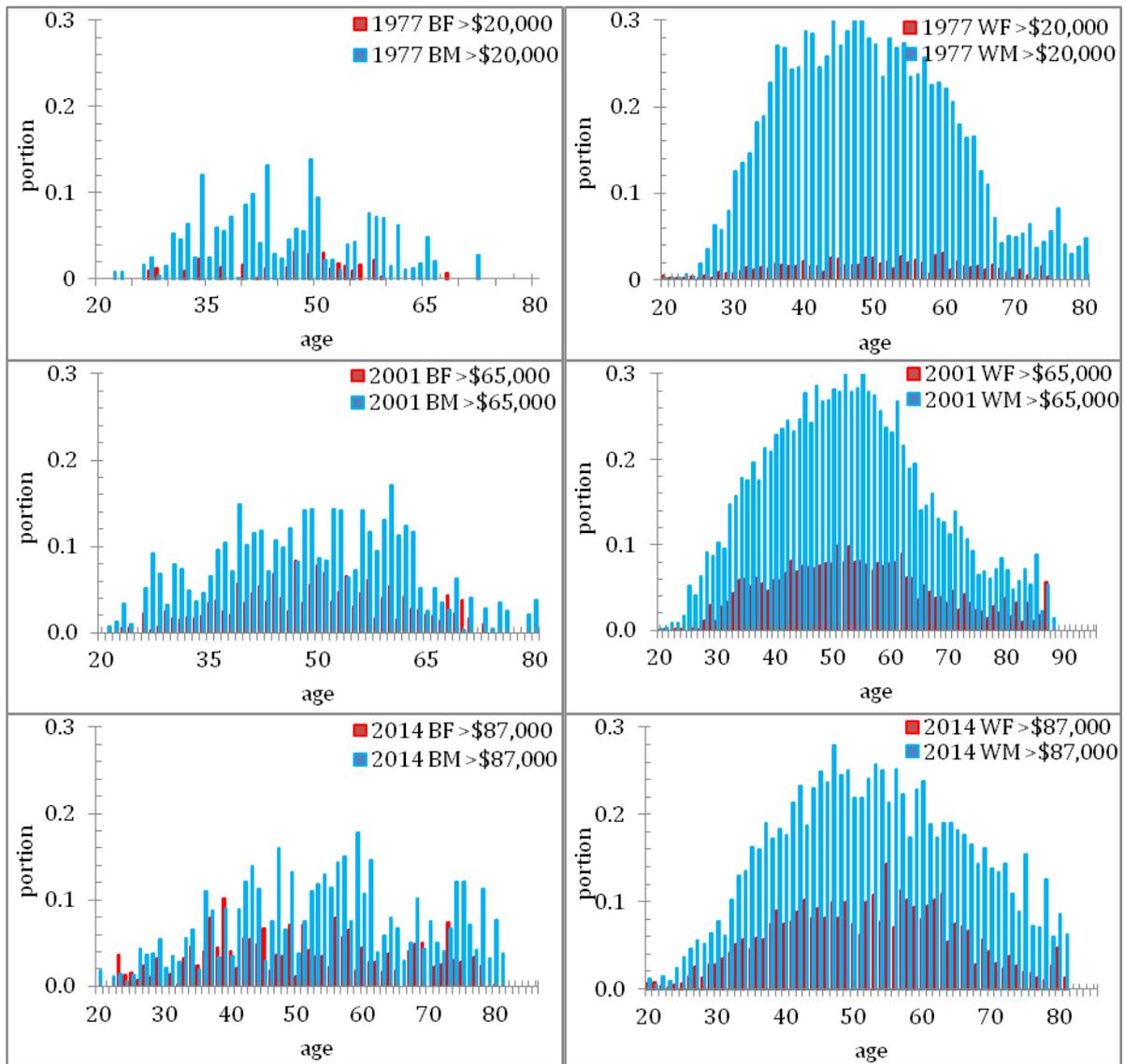

Figure 12. The age-dependent portion of people above the Pareto threshold for selected years between1962 and 2014. Thresholds are the same for all gender-race configurations for a given year. Left panel: black males and females. Right panel: white males and females.

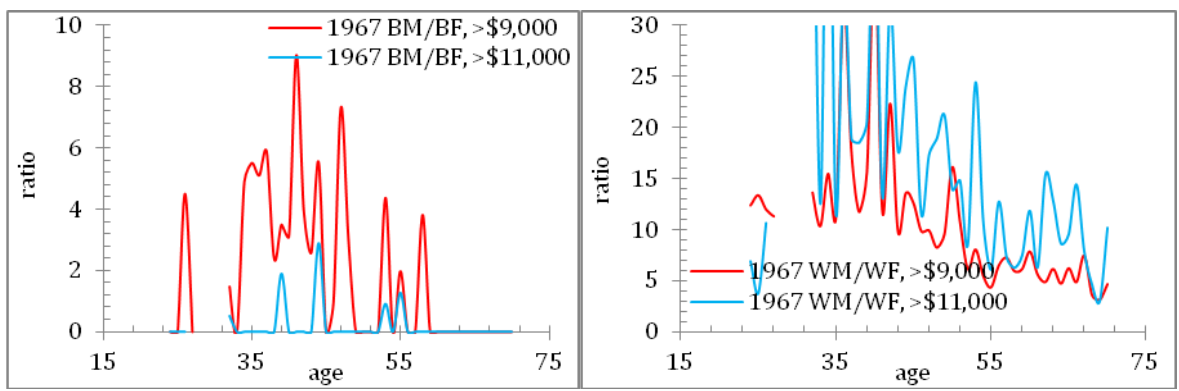

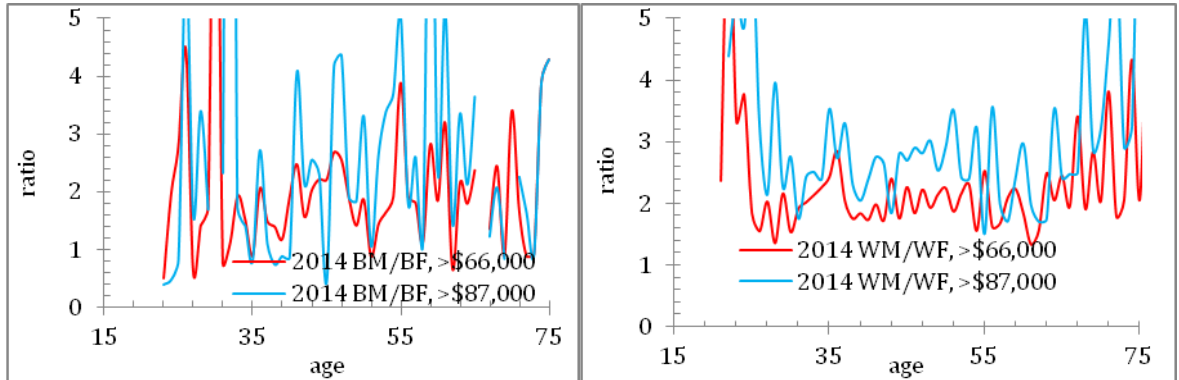
Figure 13. The male-to-female ratio for black (left panel) and white (right panel) population in 1967 and 2014. Two thresholds are shown.

However, there is no reason to think that these people have lost their capabilities. Instead of the highest instruments (work capitals) they are forced to use some smaller instruments. This replacement has two outcomes. Firstly, the largest instruments are now occupied by white males with slightly lower capabilities. This affects the rate of income growth for these "substitutes" and they reach the Pareto distribution slightly later than it would happen for people from the deprived gender and races with the highest capabilities. This effect is hard to reveal using the crude measurements provided by the CB. The CPS survey does not cover well the not-white-male population with the highest incomes. Moreover, we do not know the exact share of people which has to be in the Pareto distribution. Despite the difficulty to accurately measure this effect we can say that it is highly important for economic growth. People with the highest income really drive the economy (at the end of the day GDI = GDP) and the earlier entrance in the highest income group of even a fraction of a percent extra people will accelerate real economic growth. Best people do drive the economy.

Secondly, those (not-white-male) people who were shifted to lower instruments have to reach their peak income slightly earlier. Our model describes this effect and shows that very capable people quickly reach the largest possible income for a given instrument size. This is similar to heating of a smaller sphere (working instrument) with the same bulk density of heating sources. The time needed to reach the highest possible temperature will decrease with decreasing radius. Overall, the gifted people should decrease the age of peak mean income. This is the effect observed in Figures 8 through 11. All female and not-white-male curves are characterized by earlier peaks. This observation is a direct indication of the uniform distribution of working capabilities among both genders and all races. We are going to model these effects quantitatively and to estimate the level of downshifting.

### Discussion
1. For a given year, the dependence of mean income on age differs noticeably between genders and races.
2. The age of peak mean income depends on gender and race and increases with increasing GDP per capita.
3. The white male distribution has the largest age of the mean income peak.
4. The shape of the age-dependent mean income curve for white females and all other gender/race configurations repeats the shape of the white-male curve observed 20 to 30 years before.

5. The shape of the overall mean income dependence on age is closer to that the white-male-curve since white males has the largest part of the total income and thus contributes much to the mean income figures.
6. Our model explains all observed features by one cause – the female population and not-white races have the same distribution of the capability to earn money and consistently low sizes of work instruments (work capital) compared to those for white men.
7. Considering the same capability to earn money for white females and other races, one can conclude that they are shifted to relatively lower work capitals by force.
8. Equal (fair) distribution of income between genders and races has not been achieved yet.
9. The relatively lower instrument sizes given to white females and other races make their representation in the income bins above the Pareto threshold to be lower than it must be.
10. In turn, this effect lowers the mean income for the same age since a relatively lower numbers of rich people occur in all age groups.
11. **The faster income growth and the earlier age peak in the Pareto distribution for white females and other races indicates that their higher capacities were applied to smaller instruments (capital) in line with the deprivation of higher instrument sizes of the female population.**

### Some policy recommendations

As a principal result of this study, we propose to develop a responsible social policy aimed at the acceleration of real economic growth. Equal opportunity for genders and races to use the largest instruments of working capital will bring immediate increase in real Gross Domestic Income, i.e. in Gross Domestic Product. The effect of this social policy will be also observed in the long-run, before the distribution of instrument sizes over genders and race becomes even. Here we do not even say about the resulting social equality which will be the outcome of income equality between races and genders. This does not imply income equality between people, however.